\documentclass[aip,apl,numerical,twocolumn,superscriptaddress]{revtex4}
\usepackage{bm,graphicx,graphics,amsmath,amssymb,bm,epsfig,color}
\usepackage{euscript,tabularx}
\usepackage{textcomp}
\usepackage{longtable}

\begin{document}
\bibliographystyle{aipsamp}

\title{Dynamics of two coupled vortices in a spin valve nanopillar excited by spin transfer torque}

\author{N. Locatelli}
\altaffiliation{Corresponding author. Electronic address: nicolas.locatelli@external.thalesgroup.com}
\affiliation{Unit\'e Mixte de Physique CNRS/Thales and Universit\'e Paris Sud 11, 1 ave. A. Fresnel, 91767 Palaiseau, France}

\author{V.V. Naletov}
\altaffiliation{Permanent address : Physics Department, Kazan State University, Kazan 420008, Russia}
\affiliation{Service de Physique de l'\'Etat Condens\'e (CNRS URA 2464), CEA Saclay, 91191 Gif-sur-Yvette, France}

\author{J. Grollier}
\affiliation{Unit\'e Mixte de Physique CNRS/Thales and Universit\'e Paris Sud 11, 1 ave. A. Fresnel, 91767 Palaiseau, France}

\author{G. de Loubens}
\affiliation{Service de Physique de l'\'Etat Condens\'e (CNRS URA 2464), CEA Saclay, 91191 Gif-sur-Yvette, France}

\author{V. Cros}
\affiliation{Unit\'e Mixte de Physique CNRS/Thales and Universit\'e Paris Sud 11, 1 ave. A. Fresnel, 91767 Palaiseau, France}

\author{C. Deranlot}
\affiliation{Unit\'e Mixte de Physique CNRS/Thales and Universit\'e Paris Sud 11, 1 ave. A. Fresnel, 91767 Palaiseau, France}

\author{C. Ulysse}
\affiliation{CNRS Phynano team, Laboratoire de Photonique et de Nanostructures, Route de Nozay, 91460 Marcoussis, France}

\author{G. Faini}
\affiliation{CNRS Phynano team, Laboratoire de Photonique et de Nanostructures, Route de Nozay, 91460 Marcoussis, France}

\author{O. Klein}
\affiliation{Service de Physique de l'\'Etat Condens\'e (CNRS URA 2464), CEA Saclay, 91191 Gif-sur-Yvette, France}

\author{A. Fert}
\affiliation{Unit\'e Mixte de Physique CNRS/Thales and Universit\'e Paris Sud 11, 1 ave. A. Fresnel, 91767 Palaiseau, France}

\date{\today}

\begin{abstract}

	We investigate the dynamics of two coupled vortices driven by spin transfer. We are able to independently control with current and perpendicular field, and to detect, the respective chiralities and polarities of the two vortices. For current densities above $J=5.7*10^7 A/cm^2$, a highly coherent signal (linewidth down to 46 kHz) can be observed, with a strong dependence on the relative polarities of the vortices. It demonstrates the interest of using coupled dynamics in order to increase the coherence of the microwave signal. Emissions exhibit a linear frequency evolution with perpendicular field, with coherence conserved even at zero magnetic field.

\end{abstract}

\maketitle

	The lowest energy mode of vortex dynamics corresponds to the gyrotropic motion of the core around its equilibrium position. This mode have been studied extensively (for a review see Ref. \cite{JoNaN_8_Guslienko_2008}), and very recent works demonstrated that it could also be stimulated by spin transfer torque \cite{PRB_80_Khvalkovskiy_2009,PRB_79_Choi_2009,PRL_100_Mistral_2008,PRL_99_Ivanov_2007}. The resulting microwave signals are characterized by narrow linewidths (about 1 MHz) \cite{NP_3_Pribiag_2007,NN_4_Ruotolo_2009} and, for structures based on magnetic tunnel junctions instead of metallic nanopillars, large output powers \cite{NC_1_Dussaux_2010}, making these spin transfer vortex oscillators good candidates for applications as nanoscale frequency synthesizers.

	The reported observation of microwave emission without any applied magnetic field raises questions about the actual sources of spin torque, in particular about the role played by the static and/or dynamic behavior of the polarizer \cite{APL_96_Khvalkovskiy_2010,PRB_80_Pribiag_2009}. Here, we intentionally design samples so that both the active layer and the polarizer layer can contain a magnetic vortex, thus leading to a more complex but interesting situation of coupled vortices. The problem of interacting vortices has been rarely treated even for a field-driven excitation \cite{APL_86_Guslienko_2005,PRB_67_Shibata_2003,NP_1_Buchanan_2005,JAP_99_Chou_2006,APL_95_Jun_2009} and never for a current induced excitation. Taking benefit of this \textit{2-vortices} configuration, our objective is to establish some selection rules for the observation of highly coherent coupled vortices in terms of their relative chiralities and polarities. To do that, we first demonstrate our capability to control independently the vortices chiralities and polarities, and to detect the resulting configuration through dc transport measurements. Consequently, we can provide some clear evidence that the microwave features associated with the coupled dynamics greatly depend upon the characteristics of each vortex, notably here their relative core polarities.

	Our samples are circular nanopillars having $200$nm diameters which have been patterned in a Cu(60nm)/Py(15nm)/Cu(10nm)/Py(4nm)/Au(25nm) stack (Py=Ni$_{80}$Fe$_{20}$). The remanent magnetic configuration is a quasi-uniform configuration in the thinner Py layer and a vortex in the thicker one. Nevertheless, the injection of a dc current produces an orthoradial Oersted field that can easily lead to the nucleation of a magnetic vortex in the second magnetic layer. For the present case, a current $I_{dc}$ above $3$ mA ensures that a vortex is present in each Py layer. In the following, we denote respectively the chirality and polarity of the vortex in the thin (thick) layer $c_{1}$ and $p_1$ ($c_{2}$ and $p_{2}$). For our convention, positive chirality is defined by Oersted field direction for positive current, and positive core polarity (``up'') is associated with the positive field direction.

\begin{figure}
  \includegraphics[width=8.5cm]{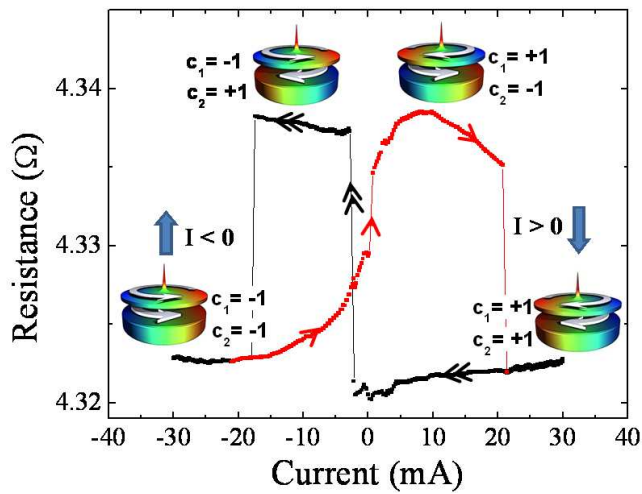}
  \caption{Resistance vs bias current at zero field for a 200nm diameter Py(15nm)/Cu(10nm)/Py(4nm) nanopillar. A parabolic contribution ($\propto I_{dc}^{2}$), due to Joule heating, has been subtracted for clarity. The sketches depict the four accessible chiralities configurations with $c_{1}$ ($c_{2}$) being the vortex chirality in the thin (thick) layer.}
\end{figure}

	In Fig. 1, we display the variation of the device resistance $R$ with $I_{dc}$ measured at zero field, which reveals that the respective vortex chiralities are independently controllable by the current through the Oersted field. At $I_{dc}= +30$~mA, the two chiralities are parallel and positive in our convention ($c_{1}=c_{2}=+1$). Upon decreasing $I_{dc}$ (follow double arrows on Fig. 1), the low resistance state is conserved until $I_{dc}=-3$~mA, even though the sign of the Oersted field is reversed when $I_{dc}$ is reversed. Below this threshold value, the Oersted field magnitude is large enough to reverse the chirality of the vortex in the thin Py layer. This transition from parallel to antiparallel chiralities is associated with a sharp jump of the resistance equal to about 80\% of the full giant magnetoresistance (GMR), measured at very low $I_{dc}$ (total GMR equal to 22 m$\Omega$). This loss of GMR is attributed to a loss of spin accumulation in the core region due to transverse spin diffusion, as described in ref. \cite{PRB_73_Urazhdin_2006}. The configuration with opposite chiralities is observed between $- 3$ and $- 18$~mA. Due to its higher stability, the vortex in the thick layer eventually switches its chirality (to $c_{2}=-1$) at this higher threshold value \cite{PRB_73_Urazhdin_2006}, resulting in a low resistance state, again associated with two vortices with parallel chiralities ($c_{1}=c_{2}=-1$). When  $I_{dc}$ is swept back from  $- 30$ to $+ 30$~mA (follow single arrows on Fig. 1), similar features corresponding to individual chirality switches are detected even though the transitions are more progressive due to the impact of the spin torque on the static configuration \cite{JAP_103_Deak_2008}. Indeed in our convention, for negative current, the spin torque destabilizes the configuration with the vortices having parallel chiralities. Thereby, in the thin layer where the vortex is less stable and the spin torque is more efficient, the vortex is distorted and then annihilated at low negative current, leading to stabilization of a quasi-uniform state around $0~mA$. The vortex is then nucleated again with inverted chirality as a positive current is applied. Similarly, spin torque destabilizes the configuration with anti-parallel chiralities for positive current, explaining the curved resistance behavior in this region.
	
\begin{figure}
  \includegraphics[width=8.5cm]{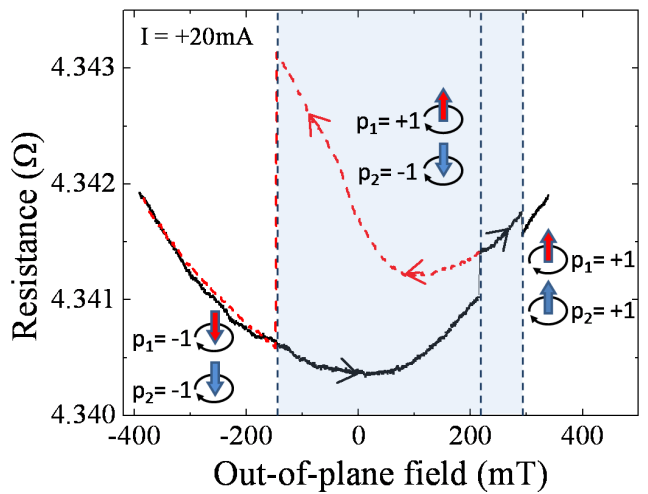}
  \caption{Resistance vs out-of-plane field $H_{perp}$ in the $c_{1}=c_{2}=+1$ configuration at $I_{dc} = +20$ mA. The solid line corresponds to the major loop for an increasing field after a negative saturation. The minor loop appears as a dashed line. The sketches represent the two vortex polarities for each configuration with $p_{1}$ ($p_{2}$) being the vortex core polarity in the thin (thick) layer. Vertical dotted lines indicate field values at which a core reversal occurs. The colored field region corresponds to the region where the highly coherent microwave emission with antiparallel core polarities can be observed.}
\end{figure}

	Then after, in order to independently control and detect the polarities of each vortex, we measure $R$ versus out-of-plane field $H_{perp}$ at $I_{dc} = + 20$~mA (see solid line in Fig 2). Note that this measurement could be achieved only in the $c_{1}=c_{2}=+1$ state at $I_{dc} > 0$ because, only there, the spin torque and the Oersted field act jointly to stabilize the two vortices' chiralities. At large negative $\mu_{0}H_{perp} $, both polarities are parallel and negative ($p_1=p_2=-1$). While increasing $\mu_{0}H_{perp}$ up to $220$ mT, a reversible variation of $R$ is measured due to a modification of the vortex shapes. Then at $\mu_{0}H_{perp} = 220$~mT, a small jump of $R$ (of about $0.4$ m$\Omega$) is detected. We relate this to the core polarity switching of the vortex in the thin Py layer, i.e. to $ p_{1}=+1$ \cite{PRB_67_Thiaville_2003,JoMaMM_240_Okuno_2002}. The two core polarities remain opposite until a larger $\mu_{0}H_{perp}$ = 280 mT is applied at which the polarity of the vortex in the thick layer eventually switches. Then $R$ returns to the low level state corresponding to parallel polarities ($p_{1}=p_{2}=+1$). The small difference of $R$, between parallel and opposite polarities, is related to the displacement of the equilibrium position of vortices' cores due to their dipolar core-core interaction. The hysteretic behavior of vortex core switching in the thin layer is demonstrated by inverting the field sweeping direction before the vortex core of the thick layer switches, resulting in the minor loop presented on Fig. 2 (see dashed line).

\begin{figure}
	\includegraphics[width=8.5cm]{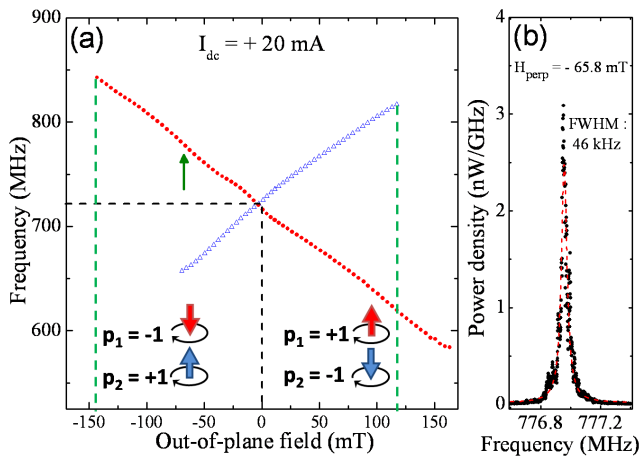}
	\caption{(a) Fundamental frequency of the coupled dynamics vs $H_{perp}$ for configurations with positive chiralities and opposite polarities (see sketches). Filled circles are obtained for $p_{1}= +1$ and $p_{2}= -1$, corresponding to the experiment described on Fig. 2, whereas open triangles are obtained for the symmetric case $p_{1}=-1$ and $p_{2}=+1$. (b) Power spectrum at $I_{dc} = + 20$~mA and $\mu_{0}H_{perp} = -65.8$~mT (see vertical arrow on Fig. 3(a)), which is characterized by one of our narrowest observed linewidths.}
\end{figure}

	Having achieved the control on vortices configuration, we then study the influence of their polarities on the dynamics of the coupled vortices. While we sweep $H_{perp}$, in the $c_{1}=c_{2}=+1$ configuration, we measure simultaneously the high frequency part of the sample voltage. An important result of our work is that a large and coherent spin-transfer-induced emission is observed \textit{solely} for one configuration of the cores, namely when the two polarities are opposite : $p_{1}p_{2}=-1$. In a previous analytical work, we demonstrated that, for a given current sign, the spin-transfer-induced precession of non-interacting vortices should only depends on their relative chiralities \cite{APL_96_Khvalkovskiy_2010}. The fact that we observe a dynamics depending on relative core polarities hence appears as a strong clue that the two vortices are interacting, leading to coupled vortex dynamics. Note that the onset of the vortices' dynamics is an other source of increase of $R$, in particular for the large increase observed in the negative field region on the minor loop in Fig 2. As can be seen on Fig. 3(a), the dependence of the emission frequency is linear with $H_{perp}$ with the sign of the slope related to the orientation of the core polarities with respect to the field direction (this results in two frequency branches as a function of $H_{perp}$). Therefore the fact that the frequency of the observed coupled mode increases when $H_{perp}$ increases in the direction of the core polarity of the thick layer ($p_2$) indicates that the dynamical behavior is governed by the vortex in the thick layer \cite{PRL_102_Loubens_2009}. In addition, by following the frequency at the crossing point of the two branches (at $H_{perp}$ = 0), as a function of $I_{dc}$, we can estimate the frequency at zero current associated to the coupled dynamic mode at $f_{0} = 563$ MHz (see Fig. 4(a)). The linear dependence of the frequency with $I_{dc}$ is consistent with the expected linear frequency increase induced by the Oersted field confinement and the spin torque \cite{PRB_80_Khvalkovskiy_2009}. This zero current frequency lies between the calculated eigen-frequencies of the two isolated layers, being a clue that a strong coupling exists between the two vortices \cite{APL_86_Guslienko_2005, PRB_67_Shibata_2003} : $f_{1} = 150$~MHz for the thin $4$~nm layer and $f_{2} = 700$~MHz for the thick $15$~nm layer ($M_{S_{1}} = 7.48 \times 10^{5}$~A/m and $M_{S_{2}} = 6.21 \times 10^{5}$~A/m, measured by mechanical ferromagnetic resonance in the nano-pillar). This emphasizes that the observed mode corresponds to one of the two modes of excitation for a pair of interacting vortices.

\begin{figure}
  \includegraphics[width=8.5cm]{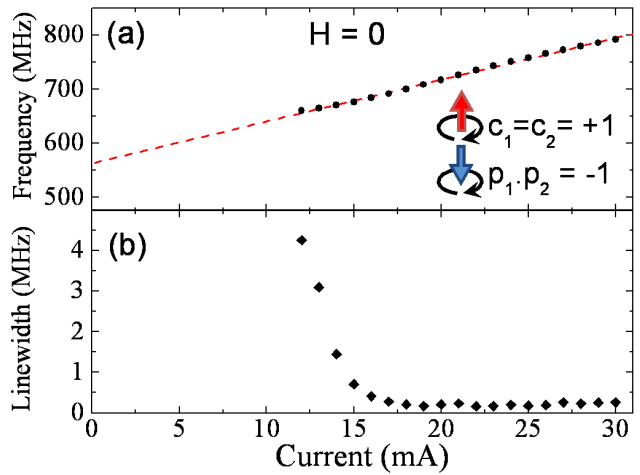}
  \caption{At $\mu_{0}H_{perp}=0$ : (a) Evolution of the zero field fundamental frequency as a function of bias current for positive chiralities and opposite polarities. The dotted line gives by extrapolation an estimate of one of the eigen frequencies of the coupled 2-vortices (b) Linewidth as a function of bias current.}
\end{figure}

	Interestingly, we note that the device emits a significant power, even at zero field, as compared to microwave emission previously observed in the absence of any applied field \cite{PRB_80_Pribiag_2009,APL_95_Devolder_2009}. The spectral linewidth, presented in Fig. 4(b), remains approximately constant for $I_{dc} > 18$ mA. Decreasing the current below this threshold value, the linewidth increases exponentially until $I_{dc} = 10$ mA where the signal associated with opposite polarities configuration disappears. However, the disappearance of the signal is not, in this case, related to a polarity reversal since the emission can always be recovered by increasing $I_{dc}$ again. Notably, a peak linewidth at zero field of only 200 kHz with a 100 pW/GHz intensity for $I_{dc} = + 20$ mA has been obtained. The most intense and narrow peak is obtained around $\mu_{0}H_{perp} = - 60$ mT, with a maximum intensity of 3 nW/GHz at the fundamental frequency together with a minimum linewidth of $46$ kHz (see Fig. 3(b)). Such an extremely narrow linewidth compared to the state of the art spin transfer nano-oscillators demonstrates the usefulness of coupled magnetic layers in tackling the important issue of spectral coherence of this kind of oscillators.

	To conclude, we demonstrate our ability, using injected current and external magnetic field, to independently control and discriminate vortex chiralities and polarities in a complex system containing two vortices, one in each magnetic layer of a trilayer GMR device. Notably, for positive current and positive chiralities, we observe a clear difference between the dc resistance levels of configurations with opposite and parallel core polarities. Peculiarly, in the dynamical regime, only configurations with $p_{1}p_{2}=-1$ are associated with emission of a high frequency voltage signal. In these configurations, the evolution of the frequency with $H_{perp}$ as well as the value of the observed frequency led us to conclude that the vortex in the thick layer plays a major role in coupled \textit{2-vortices} dynamics. The narrow linewidths that are reached, even at zero field, demonstrate the high potential of using coupled dynamics of vortices to increase the quality factor of spin transfer oscillators.

The authors acknowledge A.V. Khvalkovskiy and K.Y. Guslienko for discussion. Financial support by the CNRS and the ANR agency (VOICE PNANO-09-P231-36) and EU grant MASTER No. NMP-FP7-212257 is acknowledged.

\end{document}